\documentclass[aps,preprint,singlecolumn]{revtex4-1}

\usepackage{color}
\usepackage{amsmath}
\usepackage{amssymb}
\usepackage{bm}
\usepackage{graphicx}% Include figure files

\begin{document}

%\title{\Large Heralded single photon sources limited by multi-photon absorption}
\title{\Large Multi-photon absorption limits to heralded \\ single photon sources}

\author{Chad A. Husko$^{1,\dagger,*}$, Alex S. Clark$^{1,\dagger,*}$, Matthew J. Collins$^{1}$, Alfredo De Rossi$^2$, Sylvain Combri\'{e}$^2$, Ga\"{e}lle Lehoucq$^2$, Isabella H. Rey$^3$, Thomas F. Krauss$^4$, Chunle Xiong$^{1}$, Benjamin J. Eggleton$^{1}$}

\address{\small $^1$Centre for Ultrahigh-bandwidth Devices for Optical Systems (CUDOS), Institute of Photonics and Optical Science (IPOS), School of Physics, University of Sydney, NSW 2006, Australia \\ $^2$ Thales Research and Technology, 1 Av. A. Fresnel, 
91767 Palaiseau, France\\ $^3$ SUPA, School of Physics and Astronomy, University of St Andrews, Fife, KY16 9SS, UK \\ $^4$ Department of Physics, University of York, YORK, YO10 5DD, UK \\ {\footnotesize $^\dagger$ These authors contributed equally to this work. \\ *Correspondence and requests for materials should be addressed to chad.husko@sydney.edu.au (C.H.) and a.clark@physics.usyd.edu.au (A.C.) } }

\maketitle

%%%%%%%%%%%%%%%%%%%%%%% begin %%%%%%%%%%%%%%%%%%%%%%%%%%%%%%

%%%%%%%%%%%%%%%%%%%%%%%%%%  body  %%%%%%%%%%%%%%%%%%%%%%%%%%
\noindent
\begin{center}
\textbf{ABSTRACT}
\end{center}

\noindent
\textbf{Single photons are of paramount importance to future quantum technologies, including quantum communication and computation. Nonlinear photonic devices using parametric processes offer a straightforward route to generating photons, however additional nonlinear processes may come into play and interfere with these sources. Here we analyse these sources in the presence of multi-photon processes for the first time. We conduct experiments in silicon and gallium indium phosphide photonic crystal waveguides which display inherently different nonlinear absorption processes, namely two-photon (TPA) and three-photon absorption (ThPA), respectively. We develop a novel model capturing these diverse effects which is in excellent quantitative agreement with measurements of brightness, coincidence-to-accidental ratio (CAR) and second-order correlation function $g^{(2)}(0)$, showing that TPA imposes an intrinsic limit on heralded single photon sources. We devise a new figure of merit, the \textit{quantum utility} (QMU), enabling direct comparison and optimisation of single photon sources.}

\clearpage
%\section{Introduction}
Photons are a key building block for future quantum technologies and are the most suitable choice for linking distant quantum systems \cite{gisin2007}. Deterministic single photon sources based on quantum dots \cite{ekimov1981} and defects in diamond \cite{prawer2008diamond} are being heavily investigated, however they are currently limited due to the inherent randomness of emission direction, large bulk optical collection setups and most often the use of cryogenic temperatures, all of which make this technology challenging to scale up. The generation of photons through probabilistic processes in bulk nonlinear materials has been a workhorse for quantum information experiments \cite{kurtsiefer2001,nagata2007,lanyon2010,politi2008,peruzzo2010}.  In recent years, integrated waveguide devices have emerged as an attractive platform for scalable quantum technologies \cite{politi2008,peruzzo2010,chen2009,collins2012,takesue2007,sharping2006,clemmen2009,harada2010,davanco2012,xiong2011,xiong2012,clark2013,xiong2010OE}. These chip-scale platforms are typically made of glasses, such as silica and chalcogenide, or semiconductors such as silicon (Si) and gallium indium phosphide (GaInP). In contrast to earlier free-space investigations, one must consider additional effects induced by the media and their role in the \textit{generation} and \textit{collection} of photons. In amorphous materials, spontaneous Raman scattered photons act as a noise source, dramatically impacting the desired quantum statistics \cite{collins2012,xiong2010OE,lin2007photon}. In crystalline semiconductor materials multi-photon absorption and accompanying free carrier effects are present at larger intensities. While the roles of multi-photon effects in the classical regime are reasonably well known \cite{huskoMultiphoton,li2011FWM,colman2010,monatReview,husko2009,combrie2009,mizrahi1989,yinAgrawal2007,fishman2011XTPA}, quantum pair generation processes in these materials have only recently been investigated \cite{takesue2007,sharping2006,clemmen2009,harada2010,davanco2012,xiong2011,xiong2012,clark2013}. Though these initial observations noted that nonlinear absorption affected the experiments, at present the nuanced roles of the multi-photon processes in the quantum regime is unclear.

    In this paper we analyse the interaction of probabilistic photon generation with multi-photon processes in both the \textit{generation} and \textit{collection} of single photons for the first time. We carry out experiments in Si \cite{xiong2011,xiong2012} and GaInP \cite{clark2013} semiconductor waveguides which display inherently different nonlinear absorption processes, namely two-photon (TPA) and three-photon absorption (ThPA), respectively. We develop a novel model capturing these effects, including slow-light enhancements in the photonic crystal periodic media \cite{xiong2011,xiong2012,clark2013,huskoMultiphoton,li2011FWM,colman2010,monatReview,husko2009,combrie2009}. The series of measurements of brightness, coincidence-to-accidental ratio (CAR) and second-order correlation function $g^{(2)}(0)$ are in excellent quantitative agreement with our theoretical model, rigorously proving this framework accurate for predicting photon generation in the presence of strong nonlinear effects. We show that TPA imposes an intrinsic limit on heralded single photon generation even in the single photon pair regime. We finally devise a new figure of merit for photon generation devices, the \textit{quantum utility} or QMU, which can be used to compare and optimise photon sources across various platforms. This will be vital to the development of multiplexed and integrated photon sources, opening new routes to scalable quantum computation and simulation with large numbers of photonic qubits.

\begin{figure}[h]
%\begin{figure}[b!]
%\centering\includegraphics[width=10cm]{Fig_1_-_NL_Theory_Schematic}
\centering\includegraphics[width=10cm]{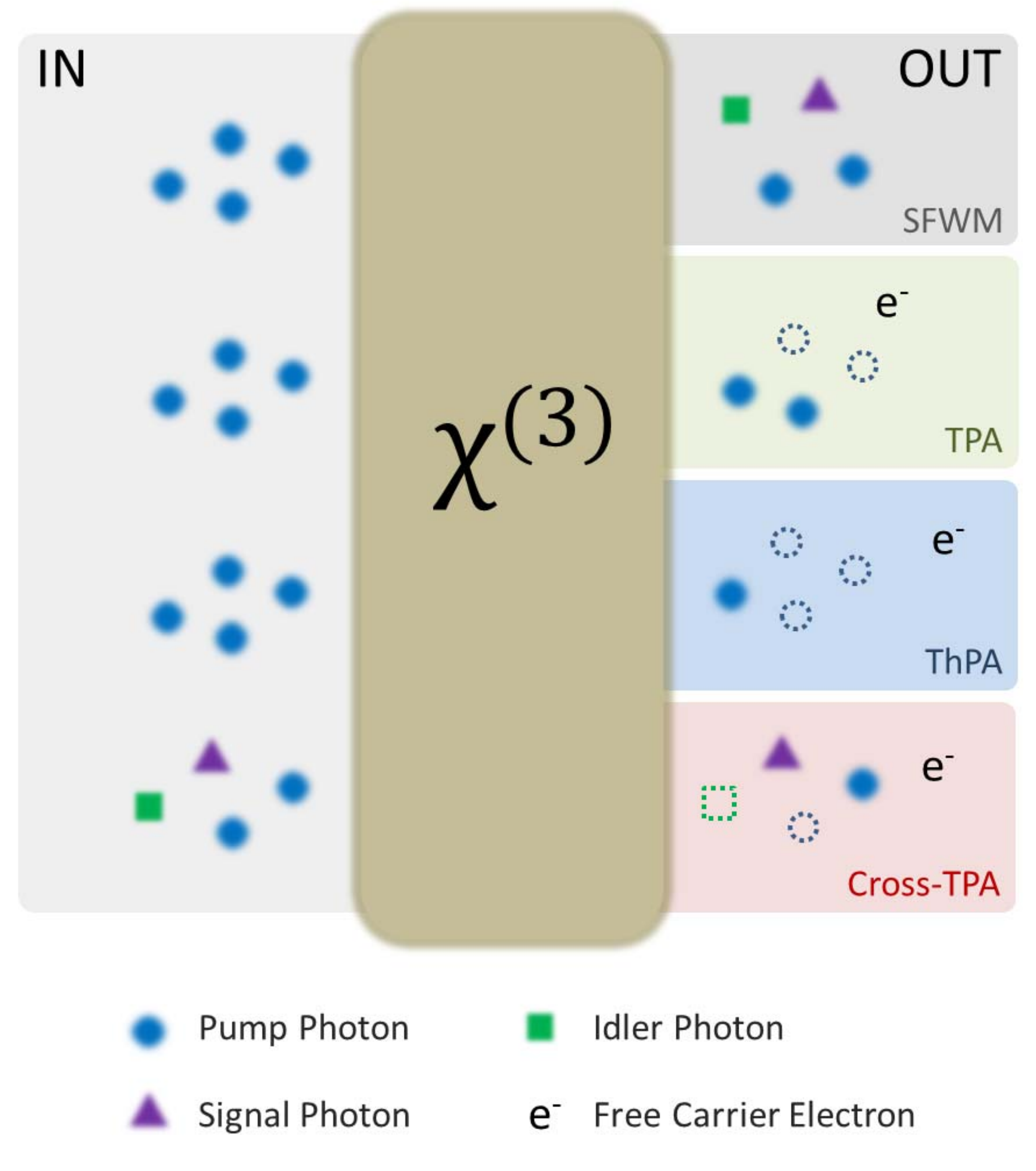}
%\vspace{0.5cm}
\caption{
\textbf{Schematic of third-order $\chi^{(3)}$ nonlinear processes.} Pump photons are input to the nonlinear material (brown). A number of processes can occur in the system including: the primary photon-pair generation process of SFWM and the secondary nonlinear processes causing loss to the pump such as TPA and ThPA, and processes causing loss of generated photons, cross-TPA (XTPA).}
%\vspace{-5mm}
\label{fig_nlProcesses}
\end{figure}

Figure \ref{fig_nlProcesses} illustrates the various nonlinear processes considered. Photons from a pulsed laser are injected to the third-order $\chi^{(3)}$ nonlinear device, where a number of processes can occur. The primary process for the generation of photons is spontaneous four-wave mixing (SFWM) where two pump photons are annihilated to generate entangled signal and idler photons at higher and lower energy, respectively. The nonlinear absorption processes include TPA, the markedly weaker ThPA, and any related free carrier absorption (FCA) induced by electrons generated from these processes. We include these losses to the pump photons propagating through the nonlinear medium, as well as through cross-TPA (XTPA) between a pump photon and a signal or idler photon. Though XTPA between the generated signal and idler photons is also possible, this is a weak effect due to the small intensities of single photons and is neglected here. In the experiments we found ThPA and cross-ThPA play negligible roles in the quantum regime. Consequently, we focus the discussion on TPA, XTPA, and FCA.
\\

%\section*{Results}
\noindent \textbf{RESULTS}
\\
\noindent\textbf{Heralded single photon generation with multi-photon absorption.} 
The novelty of our approach is the incorporation of nonlinear absorption processes in four major aspects of SFWM photon pair \textit{generation} and \textit{collection}. The \textit{generation} terms are described by the interaction of multi-photon absorption with the strong pump beam. In this regard, TPA impacts both (i) the nonlinear phase shift $\Phi$ (radians), as well as (ii) the phase matching of the parametric process. Once generated, multi-photon absorption also affects (iii) the absorption of the signal or idler single photons with the co-propagating pump beam via XTPA and (iv) absorption of single photons by free-carriers generated by the pump (FCA).

First, we consider the impact of TPA on the probability of generating a pair of photons per laser pulse, $\mu$. In the case of SFWM, this probability scales as $\mu \sim \Phi^2 \Theta^2$ with $\Phi$ and phase matching terms $\Theta$ (see Supplementary Information). In the presence of TPA, $\Phi$ scales logarithmically with power \cite{yinAgrawal2007}
\begin{equation}
\Phi=\frac{\gamma A_{eff}}{\alpha_{2}}ln\left[1+\alpha_{2} I_o S^2 L_{eff}\right],
\end{equation}

\noindent 
where $\gamma=k_o n_2/A_{eff}$ is the nonlinear parameter, Kerr coefficient $n_2$, effective area $A_{eff}$, wave number $k_0$, $\alpha_{2}$ the TPA coefficient, $I_0 = P_o/A_{eff}$ the pump intensity with peak power $P_0$, $L_{eff}=1-exp(-\alpha S L )/ (\alpha S)$ the effective device length with scattering loss $\alpha$, and $S=(n_g/n_0)$ the slow-down factor in the photonic crystal waveguide (PhCW) \cite{monatReview,li2011FWM,santagiustina2010}. Similar scaling occurs for coupled-cavity waveguides \cite{matsuda2013,davanco2012}. In the experiments we control $\Phi$ by varying the input laser power. While earlier work suggests the role of TPA in reducing coincidence counts in the SFWM process \cite{takesue2007,sharping2006,clemmen2009,harada2010,davanco2012,xiong2011,xiong2012,clark2013}, here we describe the various roles of nonlinear absorption and quantitatively include these effects in our photon pair model. While this description of $\Phi$ is pertinent to Si, we note that a linear definition of $\Phi  = \gamma P_0 L_{eff} S^2$ is appropriate for GaInP at the power levels used in these experiments. Second, as SFWM is a parametric process one must also consider the phase matching conditions. In particular, $\Theta$ is a function of $\Phi$ (see Supplementary Information). 

Further to these two mechanisms where TPA influences the \textit{pump} during pair generation, multi-photon absorption also strongly affects the generated \textit{single photons} by introducing two loss channels in the \textit{collection}. The third impact of multi-photon absorption is a collection loss due to absorption of single photons through XTPA of the signal or idler single photons with the co-propagating pump beam. This effect is described by 
\begin{equation}
\eta_{XTPA}=\frac{1}{(1+\alpha_{2}I_o L_{eff}S^2)^2},
\label{eqn:XTPA}
\end{equation}
where $\eta_{XTPA}<$1 is the collection efficiency per photon. The fourth impact of multi-photon absorption is the generation of free carriers during pump propagation via TPA. FCA of single photons is described by
\begin{equation}
\eta_{FCA}=\frac{1}{[1+\sigma N_c L_{eff}(2\alpha)]^{1/2}},
\label{eqn:FCA}
\end{equation}
\noindent where $\sigma$ is the absorption cross section, $N_c$ is the carrier density and $L_{eff}(2\alpha)=1-exp(-2\alpha L)/ (2\alpha)$ \cite{yinAgrawal2007,huskoMultiphoton}. Similar expressions exist for ThPA, though we exclude these here due to the fact that they are negligible at intensity values required for heralded single photons from SFWM. This is the first time the role of multiphoton absorption, in particular TPA, XTPA, and FCA, have been quantitatively analysed in photon pair generation and collection.

% ================================================
% ================================================
% ================================================
\clearpage
\noindent
\textbf{Experiments and modeling of SFWM pair generation in photonic crystal waveguides: linear and TPA regimes.}
We performed a number of experiments on photonic crystal waveguides fabricated in two different nonlinear semiconductor materials, namely Si and GaInP, to examine the heralded photon generation in the presence of nonlinear absorption and compare them to theoretical predictions (see Methods section for device and experimental details). At large intensities these materials exhibit two and three-photon absorption, respectively, for the photon energies (0.8 eV) in our experiments. 

\begin{figure}[h]
%\begin{figure}[b!]
%\centering\includegraphics[width=13cm]{Fig_2_-_Setup}
\centering\includegraphics[width=13cm]{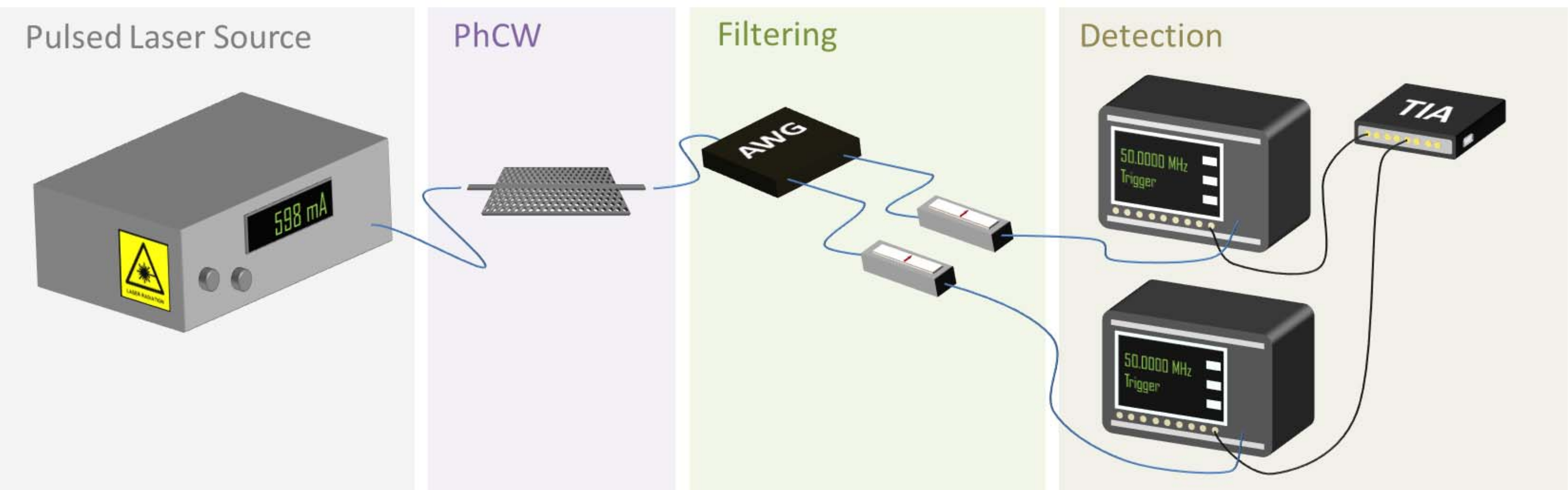}
\vspace{0.5cm}
\caption{
\textbf{Schematic of the experimental system under consideration.} A fiber laser generates 10 ps pulses in the telecommunications band at a repetition rate of 50 MHz which are injected into the photonic crystal waveguide (PhCW) under test using lensed fibers.  Photons are generated through SFWM, separated using an arrayed waveguide grating (AWG) and filtered using tunable bandpass filters.  The generated photons are then detected using single photon detector modules triggered in synchrony with the laser source and coincidence events are measured using a time interval analyser (TIA).}
%\vspace{-5mm}
\label{fig_expSchematic}
\end{figure}

To confirm these observations we create a model for the generation, filtering and detection system, shown in Fig. \ref{fig_expSchematic}, which we treat in a similar way to Bussier\'es \textit{et al.} \cite{tittel2008OE}. For the generation term, we define $p_{i}$ as a thermal distribution dependent on $\mu$, the number of pairs per pulse (including our TPA and slow-light modifications). This distribution is appropriate when the filtering matches the pump pulse bandwidth \cite{takesue2010effects}. For the photon pair collection, we create a vector describing the state of the detector system indicating the probability of each detector switching from the \textit{unclicked} state to the \textit{clicked} state (see Methods and Supplementary Information). We construct accompanying matrices describing the (i) detector dark counts $\textbf{M}_{D}$ and (ii) loss mechanisms of generated photons in the system $\textbf{M}_{\eta}$ (see Methods section). The $\textbf{M}_{\eta}$ matrix describes the photon collection efficiencies $\eta_{couple_{S,I}}$, as well as intrinsic losses of single photons due to propagation $\eta_{\alpha}=exp(-\alpha L)$ and our newly derived nonlinear loss terms $\eta_{NL} = \eta_{XTPA}\eta_{FCA}$

\begin{equation}
\eta_{S,I}=\eta_{couple_{S,I}}\eta_{\alpha}\eta_{NL},
\label{eqn:collection}
\end{equation} 
\noindent where the subscripts $S,I$ correspond to the signal and idler channels, respectively. These matrices act on an original vector set to an initial \textit{unclicked} state and are populated with a thermal distribution of photons related to $\mu$. We account for the probability of generating multiple pairs per pulse in the SFWM process through  terms proportional to $\mu^n$, where $n$ is the number of generated photon pairs. The final output vector provides the total probability for all detector states and coincidence terms.

\begin{figure}[h]
%\begin{figure}[b!]
%\centering\includegraphics[width=13cm]{Fig_3_-_Counts_and_CAR_vertical_labelled} %PDF/PNG
\centering\includegraphics[width=12cm]{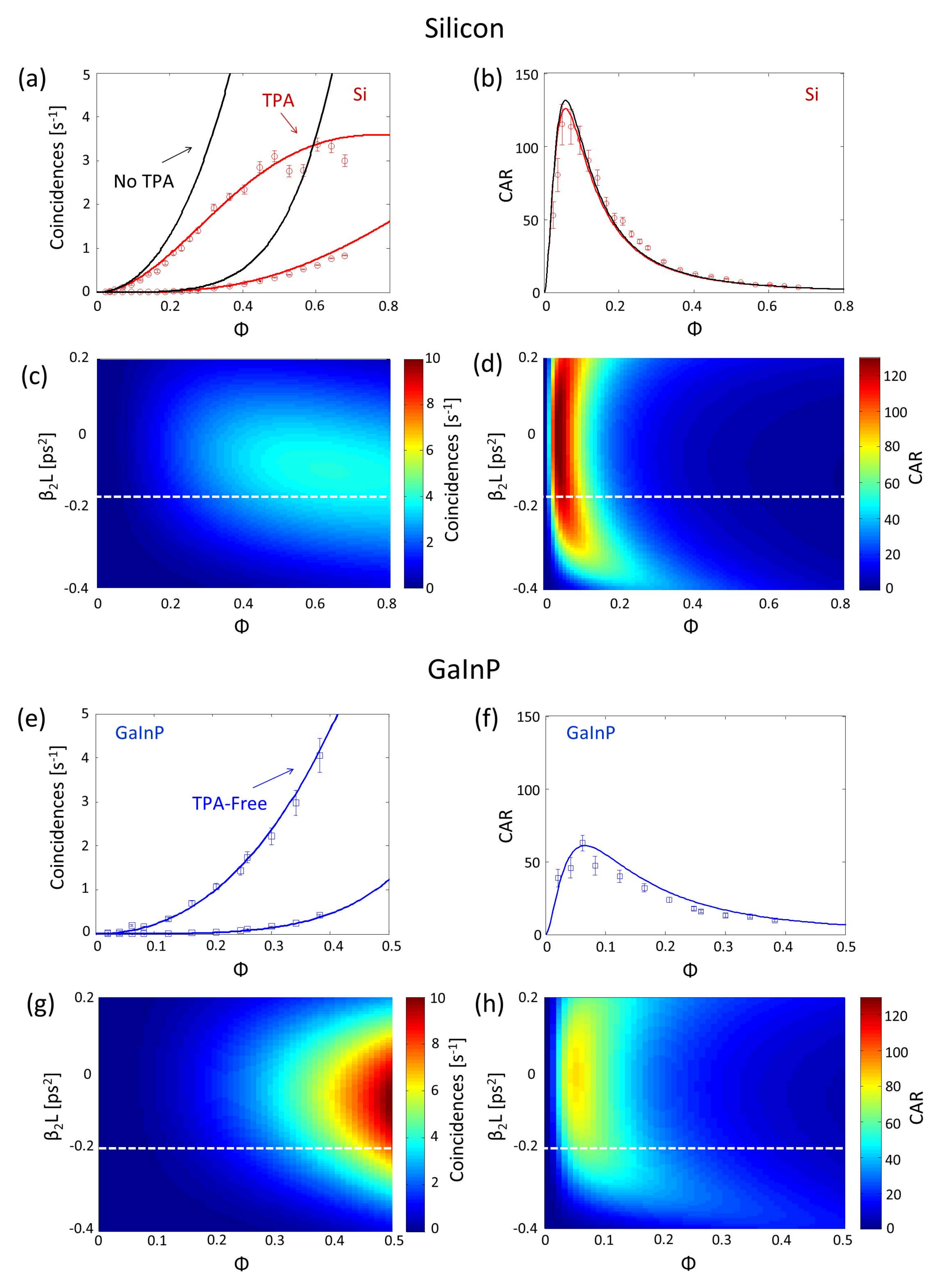} %EPS
\vspace{-0.5cm}
\caption{
\textbf{Evolution of coincidence counts and CAR plots with respect to nonlinear phase shift, $\Phi$.} (a)-(d) Plots related to Si. (a) Experimental coincidence count data with model (solid red lines) for Si. Modelling with numerically suppressed TPA (black line) is also indicated. (b) Experimental CAR versus $\Phi$ for Si. (c) Contour plots of coincidence counts and (d) CAR with respect to $\Phi$ as the GVD is varied for silicon and GaInP with fixed experimental lengths. The experimental $\beta_2 L$ is marked by the white dashed line. (e)-(h) Plots related to GaInP. GaInP shows a quadratic increase in coincidence counts due to the lack of nonlinear absorption in GaInP in the photon pair regime. All errors are calculated from Poissonian statistics.
}
%\vspace{-5mm}
\label{fig_CARexp}
\end{figure}

   Figure \ref{fig_CARexp}(a) shows the experimentally measured coincidence counts versus $\Phi$ for Si ($\beta_2$ = -0.9 ps$^2$/mm). The modelled results are in excellent agreement with the experimental data points in terms of both the SFWM photon coincidence counts (upper curves) and the accidental coincidence counts arising from detector dark counts, multi-pair emission and any residual leaked pump (lower curves). The black lines are theoretical curves with numerically suppressed TPA (and therefore FCA) showing a significant increase in both coincidences and accidentals. Of particular interest is the flattening to a maximum count of less than five-per-second for the Si device due to both the restricted $\Phi$ in the generation, as well as XTPA and FCA of the generated pairs in the collection. 

   Using the coincidence count data, we calculate the coincidence-to-accidental ratio (CAR), a metric analogous to a signal-to-noise ratio in a classical system. Fig. \ref{fig_CARexp}(b) shows the experimental data along with the modelled values once again demonstrating excellent agreement. The black curve in Fig. \ref{fig_CARexp}(b) shows a numerical suppression of TPA only slightly reducing the maximum CAR value, despite the strongly suppressed coincidence counts seen in Fig. \ref{fig_CARexp}(a). Comparing the strongly modulated coincidence counts in Fig. \ref{fig_CARexp}(a) with the relatively unaffected CAR in Fig. \ref{fig_CARexp}(b), it is clear that one must look beyond CAR as a simple metric for probabilistic photon sources.

    We calculated the coincidence counts and CAR as a function of both group velocity dispersion (GVD) and $\Phi$, including slow-light effects in the PhCW \cite{li2011FWM,monatReview,huskoMultiphoton}. Figures \ref{fig_CARexp}(c) and (d) show numerically computed contour plots of the coincidence counts for Si. The experimental device GVD-length product is indicated by the dashed white line through the plot ($\beta_2 L$ = -0.18~ps$^2$). The experimental wavelength is determined by the frequency spacing of our arrayed waveguide grating (AWG) filter. We observe in Figs. \ref{fig_CARexp}(c) and (d) that both GVD and $\Phi$ determine the best operating point, a fact that must be taken into consideration when designing and fabricating photon pair sources based on SFWM. The tilt of the coincidence contours towards anomalous (negative) GVD is required to balance the nonlinear (positive) contribution to the phase matching. We note the nonlinear contribution dominates in these chip-scale devices in contrast to the linear dispersion contribution that is more significant for fibre-based sources \cite{lin2007photon,li2004all,fan2005,clark2011}. In the case of Si, TPA additionally saturates $\Phi$, as noted previously, causing an adjustment to the nonlinear phase matching contribution for photon-pair generation. 

   We next performed identical measurements in a GaInP PhCW, a material free from TPA in the telecommunications band ($E_g$ = 1.9 eV) \cite{colman2010,combrie2009,huskoMultiphoton}. Figure \ref{fig_CARexp}(e) shows the experimentally measured coincidence counts versus $\Phi$ for GaInP ($\beta_2$ = -0.13 ps$^2$/mm). Note the x-scale ($\Phi$) is smaller for ease of comparison to the coincidence counts (y-scale) with the Si case. In contrast to the roll-off in the TPA-limited material, GaInP exhibits no deviation from the expected quadratic photon pair coincidence rate. Importantly, the ThPA intensity threshold is not reached at the power levels required to generate photon pairs. Essentially GaInP behaves as a linear material in the regimes of interest. Figure \ref{fig_CARexp}(f) indicates the measured CAR for the GaInP experiments. The maximum CAR for this device is smaller than Si due to increased linear propagation loss in the longer device of 1.5 mm versus the 0.2 mm for Si. Contour plots of the coincidences and CAR are shown in Fig. \ref{fig_CARexp}(g) and (h) for GaInP, with white dashed lines once again indicating the experimental GVD-length product ($\beta_2 L$ = -0.20~ps$^2$). Notice that despite very different behaviour in coincidence counts, the CAR plots for Si - Fig. \ref{fig_CARexp}(d) and GaInP - Fig. \ref{fig_CARexp}(h) behave comparatively at large $\Phi$ due to multi-pair generation.
\\

% ================================================
% ================================================
% ================================================

\noindent
\textbf{Second-order correlation function measurements.}
The second-order correlation function, or $g^{(2)}(0)$ is another important metric for single photon sources. The $g^{(2)}(0)$ is an indicator of how well the source output compares to a single photon \cite{clark2013,tittel2008OE,beck2007}. A $g^{(2)}(0)$ of 0 corresponds to an ideal single photon source while a standard coherent Poissonian light source exhibits a $g^{(2)}(0)$ value of 1. For our three-detector heralded measurement this is described by
\begin{equation}
g^{(2)}(0)=\frac{\Pi_{AB|H}~\Pi_{H}}{\Pi_{A|H}~\Pi_{B|H}},
\end{equation}

\noindent
where $\Pi_{j|H}$ are the conditional probabilities that the $j^{th}$ detector is triggered when the herald detector $H$ is also triggered (see Supplementary Information). We desire large $\Pi_{A|H}$ or $\Pi_{B|H}$ which indicate true single photon pairs while simultaneously minimizing  $\Pi_{AB|H}$ (triples - multiple photons) \cite{beck2007,tittel2008OE}. To measure this quantity one detects a first photon (the signal) heralding the arrival of a second photon (the idler) with this latter photon passing through a 50:50 coupler and sent to one of two detectors. True single photons will only trigger one of these two detectors. If there is a coincident count across these two detectors the $g^{(2)}(0)$ rises above the ideal value of zero indicating more than a single photon is present. 

\begin{figure}[h]
%\begin{figure}[b!]
%\centering\includegraphics[width=9cm]{Fig_4_-_g2}
\centering\includegraphics[width=9cm]{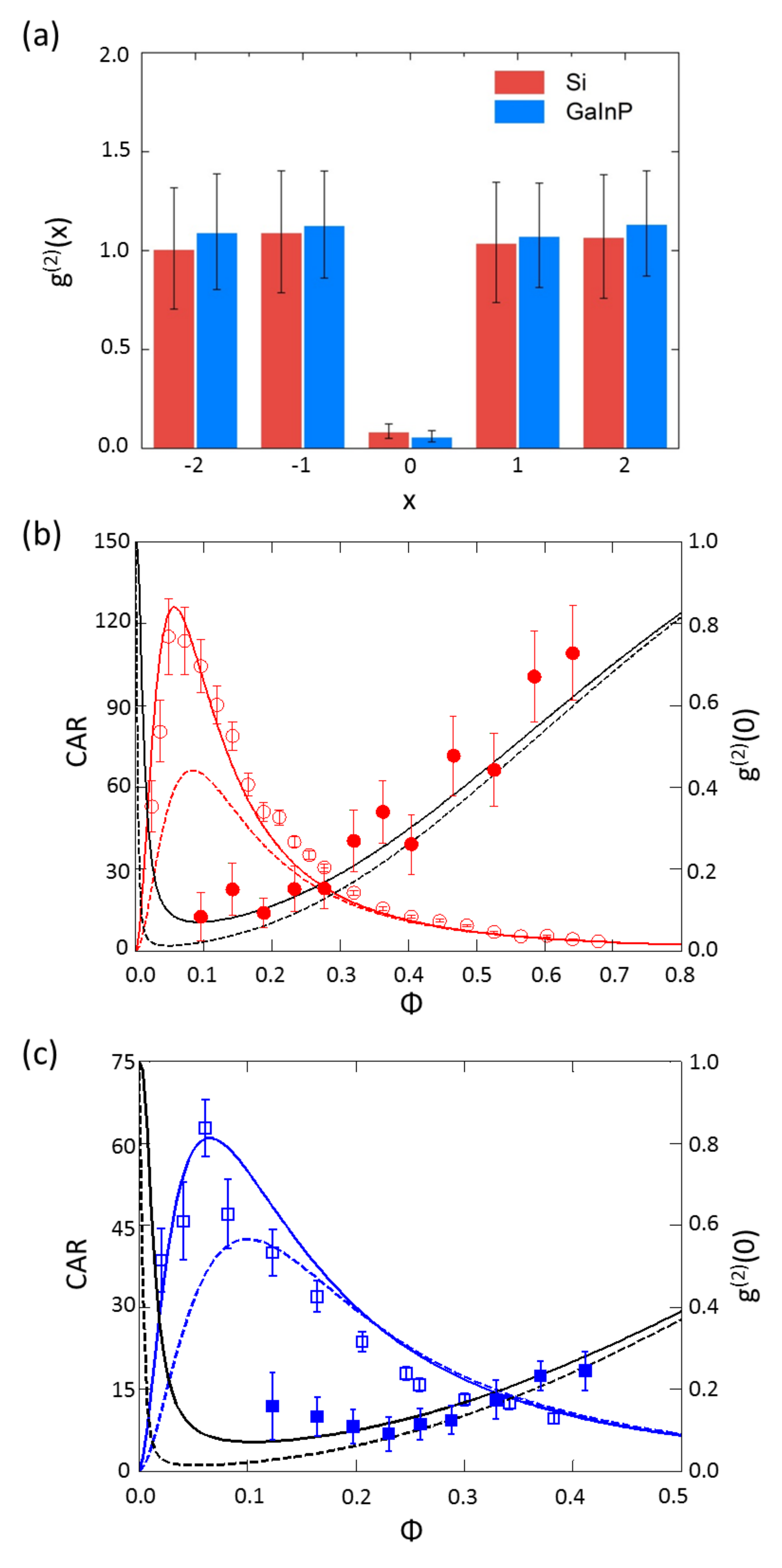}
\vspace{-0.5cm}
\caption{
\textbf{Second-order correlation function measurement results.} (a) Both Si and GaInP show clear non-classical results at zero delay with values of 0.09 and 0.06 respectively, recovering classical Poissonian statistics at alternative delays.  (b)-(c) We plot the experimental and modelled $g^{(2)}(0)$ (solid symbols and black lines) in parallel with the CAR (empty symbols and red and blue lines) for Si  (b) and GaInP (c) versus the $\Phi$. Dotted lines are numerical extrapolations for the $g^{(2)}(0)$ at the same detector settings as the CAR measurement (black dashed) and the CAR at the same detector settings as the $g^{(2)}(0)$ measurements (blue and red dashed).}
%\vspace{-5mm}
\label{fig_g20exp}
\end{figure}

     Figure \ref{fig_g20exp}(a) shows our $g^{(2)}(x)$ measurements for the two platforms, where $x$ corresponds to the integer number of delayed pulse periods of one of the heralded photons with respect to the other. True heralded single photon operation only occurs in the $g^{(2)}(0)$ window, achieving minimum $g^{(2)}(0)$ values of 0.09 for Si and 0.06 for GaInP. When $x$ is non-zero, the heralding of the output photon is lost, and the statistics revert to a coherent Poissonian source with $g^{(2)}(x)$ values of 1, as expected. 

   The evolution of the $g^{(2)}(0)$ parameter as a function of $\Phi$ is shown in Figs. \ref{fig_g20exp}(b) and (c) for Si and GaInP, respectively. To model $g^{(2)}(0)$, we extend our previous two-photon formulation to accommodate three detectors. The chief sources increasing $g^{(2)}(0)$ include dark counts at small $\Phi$ and multi-pair generation at larger $\Phi$. In both cases the extended $g^{(2)}(0)$ model is once again in solid agreement with the experimental data. The minimum of the $g^{(2)}(0)$ function corresponds to the onset of the multiple pairs-per-pulse (i$>$1) (see Supplementary Information).

   The present limit to these values is detector dark counts, not the devices themselves. In particular, for the $g^{(2)}(0)$ experiments, we required a larger detection efficiency of $\eta_D$ = 10\% (compared to 7.5\% for the CAR experiments) in order to obtain significant counting statistics for the three-photon $g^{(2)}(0)$ measurements. A unique aspect of our model is the ability to numerically extract the $g^{(2)}(0)$ value from our low dark count experimental CAR data. The black dashed lines in Figs. \ref{fig_g20exp}(b) and (c) show the extrapolated results reveal a striking $g^{(2)}(0)$ of 0.013 for GaInP and less than 0.007 in the case of Si, a near order of magnitude improvement over our prior record for PhCW devices \cite{clark2013}. We expect near ideal single-photon operation, $g^{(2)}(0)$ $\rightarrow$ 0, in our devices with ultra-low dark count detectors such as superconducting detectors \cite{divochiy2008superconducting,hadfield2009single}. The GaInP is slightly higher due to higher propagation loss in the longer device.

Conversely, we extrapolate a CAR from our experimental $g^{(2)}(0)$, shown as the red/blue dashed lines in Figs. \ref{fig_g20exp}(b) and (c). We observe the numerically extrapolated CAR is below the level achieved in the experimental CAR measurements, as expected from the higher dark count rate at $\eta_D$ = 10\%. An interesting feature of the overlaid $g^{(2)}(0)$ and CAR plots is the noticeable correspondence between the minima in the $g^{(2)}(0)$ and the maximum of the CAR. This feature is due to multi-pair generation at large $\Phi$ acting as the dominant noise source in these metrics. The $g^{(2)}(0)$ rises above a value of 0.5 for larger $\Phi$ as expected (see Supplementary Information). 
\\

\begin{figure}[h]
%\centering\includegraphics[width=12cm]{Fig_5_-_NL_losses_no_pump_small_with_legend}	%PDF
%\centering\includegraphics[width=10cm]{Fig_5_-_NL_Losses.eps}							%EPS
\centering\includegraphics[width=10cm]{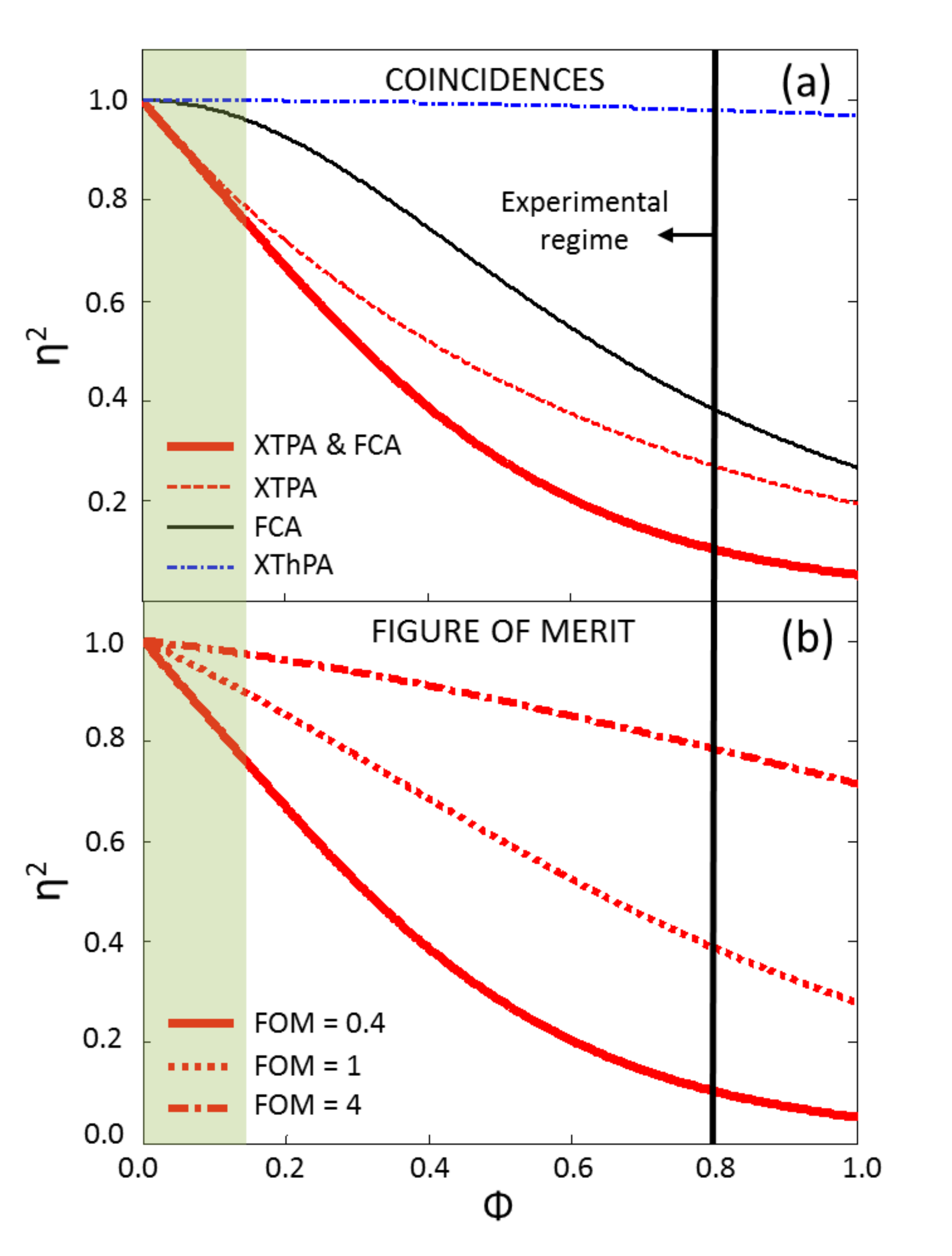}							
\caption{\textbf{Role of nonlinear loss processes.} (a) Plot of the collection efficiencies squared (proportional to coincidences) for XTPA, FCA, XThPA versus $\Phi$. It is clear that XTPA reduces the single photon count even in the single photon regime (shaded green). (b) Plot demonstrating an improved nonlinear figure-of-merit (reduced TPA) improves the photon collection efficiency. Nonetheless, photons are always lost when XTPA is present.
}
%\vspace{5mm}
%\lfig{nlLoss}
\label{fig_nlLoss}
\end{figure}

% ================================================
% ================================================
% ================================================
\noindent
\textbf{Implications of TPA and XTPA on heralded single photons.} We now look more closely into the role of the multiphoton losses in heralded single photon sources and derive some general conclusions about TPA-limited processes. We describe the photon \textit{generation}, which affects the pump beam phase shift $\Phi$ through TPA and the free carrier population $N_c$, detailed in the Supplementary Information. Here we focus on XTPA in the \textit{collection} of single photons and coincidence counts as it has a more pronounced impact than TPA on pair generation. Recall the collection losses given by Eqn. \ref{eqn:collection} are per photon channel and thus coincidence counts are proportional to the intrinsic collection losses as: $\eta^2 = \eta_\alpha^2 \eta_{XTPA}^2 \eta_{FCA}^2$. Figure \ref{fig_nlLoss}(a) shows the various collection efficiencies squared (proportional to coincidences) versus $\Phi$. We indicate the experimental region in the plot ($\Phi<$ 0.8 radians - black line), as well as the single photon pair regime shown as the shaded area. The single photon pair region was determined by setting $i = 1$ in the model and is consistent for both Si and GaInP. Immediately one notices XTPA is the dominant source of loss in the system, quickly decreasing from 100\% \textit{even in the single photon pair regime}.

The implications of this result are absolutely critical to all single photon sources exhibiting TPA. In particular, heralded single photon sources based on TPA-limited materials will require additional multiplexing, and therefore waveguides and footprint, to achieve the same efficiency compared to linear materials \cite{migdall2002,ma2011multiplex}. FCA also sets in and further reduces coincidence counts in the single pair regime and plays a larger role with increasing $\Phi$. Even at the maximum $\Phi$, the estimated contributions of XThPA is $<$0.2 dB, nearly an order of magnitude smaller than XTPA and FCA-induced pair loss at the same $\Phi$.

Finally, we consider materials with improved nonlinear figures of merit $FOM_{NL} = n_2 / (\alpha_2 \lambda)$ to discern whether TPA is truly an ultimate limit to photon pair generation. Fig. \ref{fig_nlLoss}(b) shows $\eta^2$ for various values of $FOM_{NL}$. Silicon ($FOM_{NL}$ = 0.4) is indicated as the solid line, with hypothetical values of four-fold $FOM_{NL}$ = 1 and ten-fold $FOM_{NL}$ = 4 lower $\alpha_2$ values suggesting drastic improvements. Though XTPA would be diminished in these materials, as $\eta <$ 1, it is an impairment one must always consider when present. In addition to our experimental demonstrations
in GaInP in this work \cite{colman2010,combrie2009,huskoMultiphoton}, other wide bandgap materials such as AlGaAs \cite{mizrahi1989,dolgaleva2011AlGaAs} and SiN \cite{levy2009nitride,helt2011quantum} are promising candidates for future single photon experiments. The advantages of a linear (TPA-free) material, even at the single photon level, will certainly be required for efficient photon generation.
\\
% ================================================
% ================================================
% ================================================

\noindent
\textbf{Optimisation of photon sources: the quantum utility.} It is clear from this study that developing the ideal photon source requires the optimisation of the output statistics across a number of metrics. For probabilistic photon sources, these include the achievable coincidence count rate, CAR and $g^{(2)}(0)$, which are difficult to optimise simultaneously in a practical environment.  While CAR and $g^{(2)}(0)$ are well established measures of the quality of probabilistic photon sources, at present no figure of merit exists to combine these critical quantities. With this in mind we have developed a new figure of merit for single photon sources which we term the \textit{quantum utility}, or QMU, defined as 

\begin{equation}
\textrm{QMU} = \mu' \times \textrm{SNR} \times [1-g^{(2)}(0)],
\label{eqn:QMU}
\end{equation}

\noindent where $\mu'$ is the single photon collection efficiency after any intrinsic device losses, SNR is the signal to noise ratio of desirable photons to accidental photons.

\begin{figure}[h]
%\begin{figure}[b!]
%\centering\includegraphics[width=12cm]{Fig_6_-_QMU}
\centering\includegraphics[width=12cm]{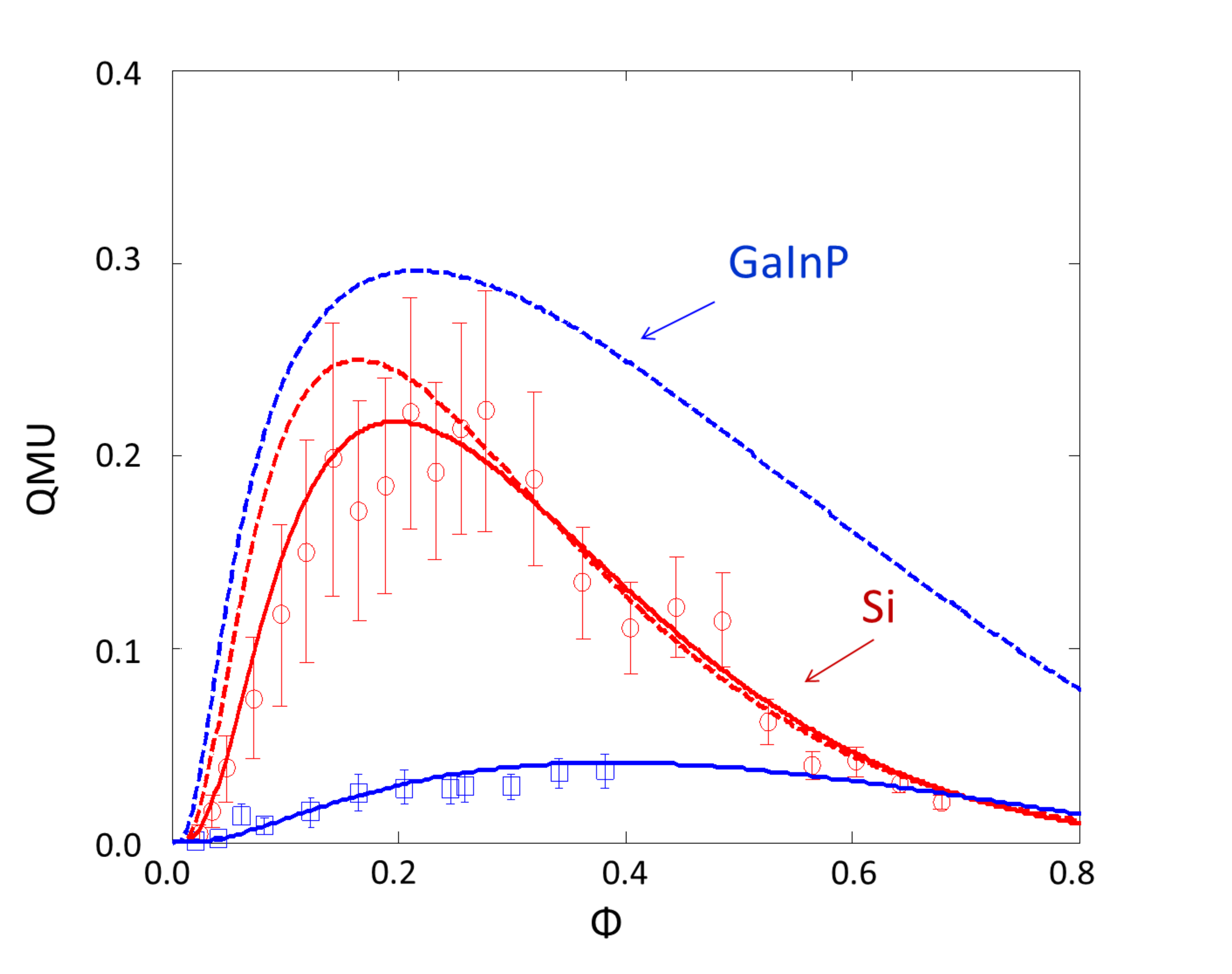}
%\centering\includegraphics[width=12cm]{Fig_6_-_QMU.eps}
%\vspace{0.5cm}
\caption{
\textbf{Quantum utility (QMU) comparison and extrapolation.} Solid lines represent the \textit{quantum utility} for the experimental devices, in excellent agreement with data for both Si (red line and circles) and GaInP (blue line and squares). Dotted lines represent a numerical fair comparison for Si and GaInP devices of similar length and effective nonlinearity operating at a more favourable dispersion. The nonlinear losses reduce the QMU for Si.}
%\vspace{-5mm}
\label{fig_QMU}
\end{figure}

For a single probabilistic source, $\mu'$ = $\mu \eta^2$, with $\eta^2$ due to propagation and nonlinear losses as defined above, and SNR $\equiv$ CAR. In the absence of dark counts and noise photons, CAR scales as 1/$\mu$, thereby making QMU reach a value of one for an ideal single probabilistic photon source. Under these conditions, the CAR and $g^{(2)}(0)$ would yield identical values, even with TPA. The experimental results, however, clearly indicate TPA, XTPA, and FCA reduce the number of coincidence counts as shown in Fig. \ref{fig_CARexp}, highlighting the importance of $\mu'$ as an indicator of useful photon pairs. Note that $\mu'$ excludes detector dark counts and any extrinsic losses. Moreover, this metric can be employed for systems without nonlinear loss by setting $\eta_{NL}$ = 1. Figure \ref{fig_QMU} shows the experimental points and modelled (solid line) values for the QMU of our devices. We see that the fabricated Si PhCW has a larger maximum QMU of $\sim$0.22. We note few other papers measure these three quantities in the same experimental setup, and it is consequently difficult to derive a QMU from the literature at present.
 
Building on our observations, we use the model to make a fair comparison of the Si and GaInP devices by assuming devices of the same length of 200 $\mu$m, same effective nonlinearities $\gamma, \alpha_2$, matched group index $n_g$=37 and coupling loss, all of which are experimentally achievable. We also consider operation at a more favourable dispersion in the current device at $\beta_2 L$ = -0.02 ps$^2$ ($\beta_2$ = -0.1 ps$^2$/mm). Fig. \ref{fig_QMU} indicates the numerically extrapolated QMU shown as dashed lines. GaInP reaches a larger QMU of 0.3 compared to Si at 0.25. In addition, the GaInP QMU is higher than Si for all $\Phi$, indicating emission of more coincident photons with a high CAR and low g$^{(2)}(0)$ over all coupled powers. Separate plots of $\mu'$, CAR, and $g^{(2)}(0)$ for the comparative case are in the Supplementary Information. At the maximum QMU for GaInP we expect a 5-fold increase in the coincidence counts whilst maintaining an excellent signal-to-noise of $\sim$ 30, meaning a four-photon experiment such as quantum interference or an integrated quantum gate could be carried out in a matter of hours rather than days.

   We close with commentary on the broader applicability of the QMU to deterministic single photon sources based on multiplexing and atom-like single photon sources \cite{ekimov1981,prawer2008diamond}. By multiplexing several of the heralded sources in this paper, it is possible to create a pseudo-deterministic single photon source \cite{migdall2002,ma2011multiplex}. Single photon heralded sources composed of multiplexed probabilistic sources is an area of active research with integrated spatial multiplexing only demonstrated this year \cite{collinsCLEOEurope} while temporal multiplexing currently remains undemonstrated \cite{mower2011timeMux}. For a multiplexed heralded source, where the multipair noise is decoupled from the heralded photon rate, a QMU $> 1$ is possible and is indeed desirable. Concretely, one can operate at the same $CAR$ across a series of multiplexed sources whilst increasing the single photon rate $\mu'$, thereby exceeding the multipair limit of a single probabilistic source. For atom-like single photon sources, QMU values greater than one are also possible when including the appropriate definitions of $\mu'$ and SNR. In these sources, $\mu'$ is the photon collection rate and SNR is the ratio of collected photons to noise photons due to dark counts and unsuppressed pump beams. With the QMU we now have a succinct metric to compare single-photon sources across multiple platforms. 
\\

% ================================================
% ================================================
% ================================================
\noindent \textbf{DISCUSSION}
%\section*{Discussion}

In summary, we have described the interaction of probabilistic photon generation with multi-photon processes in both the \textit{generation} and \textit{collection} of single photons for the first time. We measured the CAR and $g^{(2)}(0)$ of photon pairs generated from spontaneous four-wave mixing in two representative materials exhibiting two-photon absorption (Si) and three-photon absorption (GaInP), respectively. We developed a model to describe photon-pair generation in highly nonlinear semiconductors incorporating these multi-photon absorption effects and associated free-carriers. The model accurately predicts output photon statistics including source brightness, coincidence-to-accidental ratio and second-order correlation function for heralded photons. Using this model, we showed that TPA imposes an intrinsic limit on heralded single photon generation even in the single photon pair regime for different nonlinear figures of merit. Finally, we formulated the \textit{quantum utility} (QMU), a new figure of merit based on these simple metrics which allows us to optimize photon source parameters for future quantum information science experiments and technology. The development of these sources will be integral to increasing the complexity of quantum photonic experiments.

\clearpage
\noindent \textbf{METHODS}
%\section*{Methods}
\\
\textbf{Basic formulation of the model.}  
For a two-photon measurement system, such as that used to measure coincidence counts and CAR, we begin by constructing a vector which describes the joint state of the two detectors $\textbf{P}= \begin{pmatrix} P_{\overline{S}\overline{I}}, P_{{S}\overline{I}}, P_{\overline{S}{I}}, P_{{S}{I}} \end{pmatrix}$, for the signal and idler detectors $S$ and $I$, where the bar corresponds to no detection.  We initialise this vector in the unclicked state such that $\textbf{P}_0=(1,0,0,0)$ and then act upon this with matrices describing the detector dark counts, 
\textbf{M$_D$}, and system losses, \textbf{M$_\eta$}, which have size 2$^n$ x 2$^n$, where $n$ is the number of detectors for a number of photon pairs generated, $i$, such that
\begin{equation}
\textbf{P}= \sum^{\infty}_{i=0} p_{i}M_{D}(M_{\eta})^{i}\textbf{P}_0.
\end{equation}
The probability distribution of photons $p_i$ depends on the number of pairs generated per pulse, $\mu$, from a thermal source such that 
\begin{equation}
p_{i} = \frac{\mu^i}{(1+\mu)^{i+1}}.
\end{equation}			
with the nonlinear absorption affecting $\mu$ as noted in the main text. Nonlinear losses $\eta_{XTPA},\eta_{FCA}$ are included as extra terms in \textbf{M$_\eta$}, however could equally be included as separate matrices.  By analysing the output vector $\textbf{P}$ on a per pulse basis we are able to extract the coincidences and singles counts, allowing a full calculation of the other metrics for analysing photon sources including the CAR.  This method can be extended to three detectors for $g^{(2)}(0)$ correlation function calculations, as described in the Supplementary Information.
\\

\noindent \textbf{Device Specifications.}
The silicon PhCW device is fabricated from a silicon on insulator wafer with a 220 nm silicon membrane layer on 2 $\mu$m of silica using electron beam lithography and reactive ion etching.  The PhCW is made by creating a triangular lattice of holes with a row missing in the $\Gamma-K$ direction with the silicon membrane suspended in air.  The waveguide is 196 $\mu$m long with inverse tapers for improved coupling and polymer waveguides for mode matching to the lensed fibres. The two rows of holes adjacent to the waveguide are laterally shifted to engineer the dispersion \cite{li2008systematic} such that the group index is 30 and is flat to within 10\% across the SFWM bandwidth used and centred at a wavelength of 1540.6 nm. The effective nonlinearity $\gamma_{eff} = \gamma S^2 =$ 4000 W$^{-1}$m$^{-1}$ includes the slow light effect with a linear propagation loss of 50 dB.cm$^{-1}$.  The GaInP device is 1.5 mm long self-standing membrane with a lattice period of 190 nm incorporating inverse tapers to allow efficient mode conversion. Dispersion engineering is achieved through shifting the rows adjacent to the waveguide in opposite directions along the waveguide \cite{colman2012shift}, achieving a group index of 20 in the slow light region with the pump centred at 1542.5 nm. The input facet of the GaInP was damaged and limited the amount of achievable phase shift. Coupling is estimated as 9.5~dB in/2.5~dB out. The effective nonlinearity in this device is $\gamma_{eff}$ of 2300 W$^{-1}$m$^{-1}$, slightly lower than the silicon device and has a linear propagation loss of 30 dB.cm$^{-1}$. 
\\

\noindent \textbf{Experimental Setup.}  A wavelength tunable C-Band fibre laser provides 10 ps pulses at a repetition rate of 50 MHz which are passed through an isolator to protect the laser.  Any residual pump cavity photons are removed by a WDM and any spontaneous Raman scattered photons from the silica fibers are removed with a tunable band-pass filter. The input power is monitored using a power meter.  The input pulses are optimized for TE polarization using a polarization controller and coupled to the PhCW under test using 2 $\mu$m spot size lensed fibres. Signal and idler photons generated in the PhCW through SFWM are then separated using an arrayed waveguide grating which provides 40 dB of isolation from the pump.  Further pump rejection is provided by tunable band-pass filters.  We select AWG channels 600 GHz from the pump channel.  The photons are then detected using InGaAs avalanche photodiode detectors (ID Quantique ID210) synchronized to the laser repetition rate and set to a detection efficiency of 7.5\% resulting in dark counts of ~1$\times$10$^{-6}$ and ~2$\times$10$^{-6}$ per detection gate in the signal and idler detectors respectively. Coincidence counts are measured using a time interval analyser (TIA). For the $g^{(2)}(x)$ correlation function measurements a 50:50 fiber coupler is added to the idler arm with the outputs connected directly to two InGaAs APDs (ID Quantique ID201) with efficiencies of 10\% and dark count levels of ~5$\times$10$^{-4}$ per detection  gate.  The efficiency of the detector in the heralding signal photon arm is increased to 10\% to increase the 3-fold coincidence rate and reduce measurement errors.  This results in a marked increase to the dark counts in the heralding detector to~4$\times$10$^{-6}$ per detection gate.

\clearpage
\noindent \textbf{Acknowledgements}

\noindent 
The authors thank Alex Judge for discussions on the CAR model as well as Luke Helt and Mike Steel for comments on the manuscript. This work was supported in part by the Centre of Excellence (CUDOS, project number CE110001018), Laureate Fellowship (FL120100029) and Discovery Early Career Researcher Award (DE120102069, DE130101148, and DE120100226) programs of the Australian Research Council (ARC), EPSRC UK Silicon Photonics (Grant reference EP/F001428/1), EU FP7 GOSPEL project (grant no. 219299), and EU FP7 COPERNICUS (grant no. 249012).
\\

\noindent \textbf{Author contributions}

\noindent
A.C., C.H. and M.C. performed the measurements on the setup developed by M.C., C.X. and A.C. C.H. and A.C. performed the data analysis and modeling. C.H. derived the nonlinear analytic expressions and numerical simulations. S.C., G.L., and I.R. prepared the samples and nanofabrication. B.E., A.D.R., and T.K. supervised the project. C.H. and A.C. wrote the manuscript. All authors confirm the advances described in the paper.
\\

\noindent \textbf{Competing Financial Interests}
\noindent The authors declare no competing financial interests.

\clearpage
%--------------------------------------------------------------------------
%--------------------------------------------------------------------------
\clearpage
%--------------------------------------------------------------------------
%--------------------------------------------------------------------------
\clearpage
%--------------------------------------------------------------------------
%--------------------------------------------------------------------------
\clearpage
%--------------------------------------------------------------------------
%--------------------------------------------------------------------------
\clearpage
%--------------------------------------------------------------------------
%\begin{figure}[!t]
%--------------------------------------------------------------------------
\clearpage
%--------------------------------------------------------------------------
%--------------------------------------------------------------------------

% %%%%%%%%%%%%%%%%%%%%%%%%%%%%%%%%%%%%%%%%%%%%%%%
%				BIBLIOGRAPHY
% %%%%%%%%%%%%%%%%%%%%%%%%%%%%%%%%%%%%%%%%%%%%%%%
\clearpage
%\bibliographystyle{osajnl}
%\bibliographystyle{naturemag}
%\bibliography{NonlinearQMpaper_mainBiblio}

\begin{thebibliography}{10}
\expandafter\ifx\csname url\endcsname\relax
  \def\url#1{\texttt{#1}}\fi
\expandafter\ifx\csname urlprefix\endcsname\relax\def\urlprefix{URL }\fi
\providecommand{\bibinfo}[2]{#2}
\providecommand{\eprint}[2][]{\url{#2}}

\bibitem{gisin2007}
\bibinfo{author}{Gisin, N.} \& \bibinfo{author}{Thew, R.}
\newblock \bibinfo{title}{Quantum communication}.
\newblock \emph{\bibinfo{journal}{Nat. Phot.}} \textbf{\bibinfo{volume}{1}},
  \bibinfo{pages}{165--171} (\bibinfo{year}{2007}).

\bibitem{ekimov1981}
\bibinfo{author}{Ekimov, A.~I.} \& \bibinfo{author}{Onushchenko, A.~A.}
\newblock \bibinfo{title}{Quantum size effect in three dimensional microscopic
  semiconductor crystals}.
\newblock \emph{\bibinfo{journal}{JETP Lett.}} \textbf{\bibinfo{volume}{34}},
  \bibinfo{pages}{345} (\bibinfo{year}{1981}).

\bibitem{prawer2008diamond}
\bibinfo{author}{Prawer, S.} \& \bibinfo{author}{Greentree, A.~D.}
\newblock \bibinfo{title}{Diamond for quantum computing}.
\newblock \emph{\bibinfo{journal}{Science}} \textbf{\bibinfo{volume}{320}},
  \bibinfo{pages}{1601--1602} (\bibinfo{year}{2008}).

\bibitem{kurtsiefer2001}
\bibinfo{author}{Kurtsiefer, C.}, \bibinfo{author}{Oberparleiter, M.} \&
  \bibinfo{author}{Weinfurter, H.}
\newblock \bibinfo{title}{High-efficiency entangled photon pair collection in
  type-ii parametric fluorescence}.
\newblock \emph{\bibinfo{journal}{Phys. Rev. A}} \textbf{\bibinfo{volume}{64}},
  \bibinfo{pages}{23802} (\bibinfo{year}{2001}).

\bibitem{nagata2007}
\bibinfo{author}{Nagata, T.}, \bibinfo{author}{Okamoto, R.},
  \bibinfo{author}{O'Brien, J.~L.}, \bibinfo{author}{Sasaki, K.} \&
  \bibinfo{author}{Takeuchi, S.}
\newblock \bibinfo{title}{Beating the standard quantum limit with
  four-entangled photons}.
\newblock \emph{\bibinfo{journal}{Science}} \textbf{\bibinfo{volume}{316}},
  \bibinfo{pages}{726--729} (\bibinfo{year}{2007}).

\bibitem{lanyon2010}
\bibinfo{author}{Lanyon, B.~P.} \emph{et~al.}
\newblock \bibinfo{title}{Towards quantum chemistry on a quantum computer}.
\newblock \emph{\bibinfo{journal}{Nat. Chem.}} \textbf{\bibinfo{volume}{2}},
  \bibinfo{pages}{106--111} (\bibinfo{year}{2010}).

\bibitem{politi2008}
\bibinfo{author}{Politi, A.}, \bibinfo{author}{Cryan, M.~J.},
  \bibinfo{author}{Rarity, J.~G.}, \bibinfo{author}{Yu, S.} \&
  \bibinfo{author}{O'Brien, J.~L.}
\newblock \bibinfo{title}{Silica-on-silicon waveguide quantum circuits}.
\newblock \emph{\bibinfo{journal}{Science}} \textbf{\bibinfo{volume}{320}},
  \bibinfo{pages}{646--649} (\bibinfo{year}{2008}).

\bibitem{peruzzo2010}
\bibinfo{author}{Peruzzo, A.} \emph{et~al.}
\newblock \bibinfo{title}{Quantum walks of correlated photons}.
\newblock \emph{\bibinfo{journal}{Science}} \textbf{\bibinfo{volume}{329}},
  \bibinfo{pages}{1500--1503} (\bibinfo{year}{2010}).

\bibitem{chen2009}
\bibinfo{author}{Chen, J.}, \bibinfo{author}{Pearlman, A.~J.},
  \bibinfo{author}{Ling, A.}, \bibinfo{author}{Fan, J.} \&
  \bibinfo{author}{Migdall, A.}
\newblock \bibinfo{title}{A versatile waveguide source of photon pairs for
  chip-scale quantum information processing}.
\newblock \emph{\bibinfo{journal}{Opt. Express}} \textbf{\bibinfo{volume}{17}},
  \bibinfo{pages}{6727--6740} (\bibinfo{year}{2009}).

\bibitem{collins2012}
\bibinfo{author}{Collins, M.~J.} \emph{et~al.}
\newblock \bibinfo{title}{Low raman-noise correlated photon-pair generation in
  a dispersion-engineered chalcogenide as$_2$s$_3$ planar waveguide}.
\newblock \emph{\bibinfo{journal}{Opt. Lett.}} \textbf{\bibinfo{volume}{37}},
  \bibinfo{pages}{3393--3395} (\bibinfo{year}{2012}).

\bibitem{takesue2007}
\bibinfo{author}{Takesue, H.} \emph{et~al.}
\newblock \bibinfo{title}{Entanglement generation using silicon wire
  waveguide}.
\newblock \emph{\bibinfo{journal}{Appl. Phys. Lett.}}
  \textbf{\bibinfo{volume}{91}}, \bibinfo{pages}{201108--201108}
  (\bibinfo{year}{2007}).

\bibitem{sharping2006}
\bibinfo{author}{Sharping, J.~E.} \emph{et~al.}
\newblock \bibinfo{title}{Generation of correlated photons in nanoscale silicon
  waveguides}.
\newblock \emph{\bibinfo{journal}{Opt. Express}} \textbf{\bibinfo{volume}{14}},
  \bibinfo{pages}{12388--12393} (\bibinfo{year}{2006}).

\bibitem{clemmen2009}
\bibinfo{author}{Clemmen, S.} \emph{et~al.}
\newblock \bibinfo{title}{Continuous wave photon pair generation in
  silicon-on-insulator waveguides and ring resonators}.
\newblock \emph{\bibinfo{journal}{Opt. Express}} \textbf{\bibinfo{volume}{17}},
  \bibinfo{pages}{16558} (\bibinfo{year}{2009}).

\bibitem{harada2010}
\bibinfo{author}{Harada, K.-I.} \emph{et~al.}
\newblock \bibinfo{title}{Frequency and polarization characteristics of
  correlated photon-pair generation using a silicon wire waveguide}.
\newblock \emph{\bibinfo{journal}{IEEE JSTQE}} \textbf{\bibinfo{volume}{16}},
  \bibinfo{pages}{325--331} (\bibinfo{year}{2010}).

\bibitem{davanco2012}
\bibinfo{author}{Davan{\c{c}}o, M.} \emph{et~al.}
\newblock \bibinfo{title}{Telecommunications-band heralded single photons from
  a silicon nanophotonic chip}.
\newblock \emph{\bibinfo{journal}{Appl. Phys. Lett.}}
  \textbf{\bibinfo{volume}{100}}, \bibinfo{pages}{261104--261104}
  (\bibinfo{year}{2012}).

\bibitem{xiong2011}
\bibinfo{author}{Xiong, C.} \emph{et~al.}
\newblock \bibinfo{title}{Slow-light enhanced correlated photon pair generation
  in a silicon photonic crystal waveguide}.
\newblock \emph{\bibinfo{journal}{Opt. Lett.}} \textbf{\bibinfo{volume}{36}},
  \bibinfo{pages}{3413--3415} (\bibinfo{year}{2011}).

\bibitem{xiong2012}
\bibinfo{author}{Xiong, C.} \emph{et~al.}
\newblock \bibinfo{title}{Characteristics of correlated photon pairs generated
  in ultracompact silicon slow-light photonic crystal waveguides}.
\newblock \emph{\bibinfo{journal}{IEEE JSTQE}} \textbf{\bibinfo{volume}{18}},
  \bibinfo{pages}{1676--1683} (\bibinfo{year}{2012}).

\bibitem{clark2013}
\bibinfo{author}{Clark, A.~S.} \emph{et~al.}
\newblock \bibinfo{title}{Heralded single-photon source in a iii--v photonic
  crystal}.
\newblock \emph{\bibinfo{journal}{Opt. Lett.}} \textbf{\bibinfo{volume}{38}},
  \bibinfo{pages}{649--651} (\bibinfo{year}{2013}).

\bibitem{xiong2010OE}
\bibinfo{author}{Xiong, C.} \emph{et~al.}
\newblock \bibinfo{title}{Quantum-correlated photon pair generation in
  chalcogenide $as_{2}s_{3}$ waveguides}.
\newblock \emph{\bibinfo{journal}{Opt. Express}} \textbf{\bibinfo{volume}{18}},
  \bibinfo{pages}{16206--16216} (\bibinfo{year}{2010}).

\bibitem{lin2007photon}
\bibinfo{author}{Lin, Q.}, \bibinfo{author}{Yaman, F.} \&
  \bibinfo{author}{Agrawal, G.~P.}
\newblock \bibinfo{title}{Photon-pair generation in optical fibers through
  four-wave mixing: Role of raman scattering and pump polarization}.
\newblock \emph{\bibinfo{journal}{Phys. Rev. A}} \textbf{\bibinfo{volume}{75}},
  \bibinfo{pages}{023803} (\bibinfo{year}{2007}).

\bibitem{huskoMultiphoton}
\bibinfo{author}{Husko, C.}, \bibinfo{author}{Colman, P.},
  \bibinfo{author}{Combri{\'e}, S.}, \bibinfo{author}{{De Rossi}, A.} \&
  \bibinfo{author}{Wong, C.~W.}
\newblock \bibinfo{title}{Effect of multiphoton absorption and free carriers in
  slow-light photonic crystal waveguides}.
\newblock \emph{\bibinfo{journal}{Opt. Lett.}} \textbf{\bibinfo{volume}{36}},
  \bibinfo{pages}{2239--2241} (\bibinfo{year}{2011}).

\bibitem{li2011FWM}
\bibinfo{author}{Li, J.}, \bibinfo{author}{O{'}Faolain, L.},
  \bibinfo{author}{Rey, I.} \& \bibinfo{author}{Krauss, T.~F.}
\newblock \bibinfo{title}{Four-wave mixing in photonic crystal waveguides: slow
  light enhancement and limitations}.
\newblock \emph{\bibinfo{journal}{Opt. Express}} \textbf{\bibinfo{volume}{19}},
  \bibinfo{pages}{4458--4463} (\bibinfo{year}{2011}).

\bibitem{colman2010}
\bibinfo{author}{Colman, P.} \emph{et~al.}
\newblock \bibinfo{title}{Temporal solitons and pulse compression in photonic
  crystal waveguides}.
\newblock \emph{\bibinfo{journal}{Nat. Phot.}} \textbf{\bibinfo{volume}{4}},
  \bibinfo{pages}{862--868} (\bibinfo{year}{2010}).

\bibitem{monatReview}
\bibinfo{author}{Monat, C.}, \bibinfo{author}{De~Sterke, M.} \&
  \bibinfo{author}{Eggleton, B.~J.}
\newblock \bibinfo{title}{Slow light enhanced nonlinear optics in periodic
  structures}.
\newblock \emph{\bibinfo{journal}{J. of Optics}} \textbf{\bibinfo{volume}{12}},
  \bibinfo{pages}{104003} (\bibinfo{year}{2010}).

\bibitem{husko2009}
\bibinfo{author}{Husko, C.} \emph{et~al.}
\newblock \bibinfo{title}{Non-trivial scaling of self-phase modulation and
  three-photon absorption in iii-v photonic crystal waveguides}.
\newblock \emph{\bibinfo{journal}{Opt. Express}} \textbf{\bibinfo{volume}{17}},
  \bibinfo{pages}{22442--22451} (\bibinfo{year}{2009}).

\bibitem{combrie2009}
\bibinfo{author}{Combri{\'e}, S.}, \bibinfo{author}{Tran, Q.~V.},
  \bibinfo{author}{{De Rossi}, A.}, \bibinfo{author}{Husko, C.} \&
  \bibinfo{author}{Colman, P.}
\newblock \bibinfo{title}{High quality gainp nonlinear photonic crystals with
  minimized nonlinear absorption}.
\newblock \emph{\bibinfo{journal}{Appl. Phys. Lett.}}
  \textbf{\bibinfo{volume}{95}}, \bibinfo{pages}{221108--221108}
  (\bibinfo{year}{2009}).

\bibitem{mizrahi1989}
\bibinfo{author}{Mizrahi, V.}, \bibinfo{author}{DeLong, K.~W.},
  \bibinfo{author}{Stegeman, G.~I.}, \bibinfo{author}{Saifi, M.~A.} \&
  \bibinfo{author}{Andrejco, M.~J.}
\newblock \bibinfo{title}{Two-photon absorption as a limitation to all-optical
  switching}.
\newblock \emph{\bibinfo{journal}{Opt. Lett.}} \textbf{\bibinfo{volume}{14}},
  \bibinfo{pages}{1140--1142} (\bibinfo{year}{1989}).

\bibitem{yinAgrawal2007}
\bibinfo{author}{Yin, L.} \& \bibinfo{author}{Agrawal, G.~P.}
\newblock \bibinfo{title}{Impact of two-photon absorption on self-phase
  modulation in silicon waveguides}.
\newblock \emph{\bibinfo{journal}{Opt. Lett.}} \textbf{\bibinfo{volume}{32}},
  \bibinfo{pages}{2031--2033} (\bibinfo{year}{2007}).

\bibitem{fishman2011XTPA}
\bibinfo{author}{Fishman, D.~A.} \emph{et~al.}
\newblock \bibinfo{title}{Sensitive mid-infrared detection in wide-bandgap
  semiconductors using extreme non-degenerate two-photon absorption}.
\newblock \emph{\bibinfo{journal}{Nat. Phot.}} \textbf{\bibinfo{volume}{5}},
  \bibinfo{pages}{561--565} (\bibinfo{year}{2011}).

\bibitem{santagiustina2010}
\bibinfo{author}{Santagiustina, M.}, \bibinfo{author}{Someda, C.},
  \bibinfo{author}{Vadala, G.}, \bibinfo{author}{Combri\'{e}, S.} \&
  \bibinfo{author}{{De Rossi}, A.}
\newblock \bibinfo{title}{Theory of slow light enhanced four-wave mixing in
  photonic crystal waveguides}.
\newblock \emph{\bibinfo{journal}{Opt. Express}} \textbf{\bibinfo{volume}{18}},
  \bibinfo{pages}{21024} (\bibinfo{year}{2010}).

\bibitem{matsuda2013}
\bibinfo{author}{Matsuda, N.} \emph{et~al.}
\newblock \bibinfo{title}{Slow light enhanced correlated photon pair generation
  in photonic-crystal coupled-resonator optical waveguides}.
\newblock \emph{\bibinfo{journal}{Opt. Express}} \textbf{\bibinfo{volume}{21}},
  \bibinfo{pages}{8596--8604} (\bibinfo{year}{2013}).

\bibitem{tittel2008OE}
\bibinfo{author}{Bussi\`eres, F.}, \bibinfo{author}{Slater, J.~A.},
  \bibinfo{author}{Godbout, N.} \& \bibinfo{author}{Tittel, W.}
\newblock \bibinfo{title}{Fast and simple characterization of a photon pair
  source}.
\newblock \emph{\bibinfo{journal}{Opt. Express}} \textbf{\bibinfo{volume}{16}},
  \bibinfo{pages}{17060--17069} (\bibinfo{year}{2008}).

\bibitem{takesue2010effects}
\bibinfo{author}{Takesue, H.} \& \bibinfo{author}{Shimizu, K.}
\newblock \bibinfo{title}{Effects of multiple pairs on visibility measurements
  of entangled photons generated by spontaneous parametric processes}.
\newblock \emph{\bibinfo{journal}{Opt. Comm.}} \textbf{\bibinfo{volume}{283}},
  \bibinfo{pages}{276--287} (\bibinfo{year}{2010}).

\bibitem{li2004all}
\bibinfo{author}{Li, X.}, \bibinfo{author}{Chen, J.}, \bibinfo{author}{Voss,
  P.}, \bibinfo{author}{Sharping, J.} \& \bibinfo{author}{Kumar, P.}
\newblock \bibinfo{title}{All-fiber photon-pair source for quantum
  communications: Improved generation of correlated photons}.
\newblock \emph{\bibinfo{journal}{Opt. Express}} \textbf{\bibinfo{volume}{12}},
  \bibinfo{pages}{3737--3744} (\bibinfo{year}{2004}).

\bibitem{fan2005}
\bibinfo{author}{Fan, J.}, \bibinfo{author}{Migdall, A.} \&
  \bibinfo{author}{Wang, L.~J.}
\newblock \bibinfo{title}{Efficient generation of correlated photon pairs in a
  microstructure fiber}.
\newblock \emph{\bibinfo{journal}{Opt. Lett.}} \textbf{\bibinfo{volume}{30}},
  \bibinfo{pages}{3368--3370} (\bibinfo{year}{2005}).

\bibitem{clark2011}
\bibinfo{author}{Clark, A.~S.} \emph{et~al.}
\newblock \bibinfo{title}{Intrinsically narrowband pair photon generation in
  microstructured fibres}.
\newblock \emph{\bibinfo{journal}{New J. Phys.}} \textbf{\bibinfo{volume}{13}},
  \bibinfo{pages}{065009} (\bibinfo{year}{2011}).

\bibitem{beck2007}
\bibinfo{author}{Beck, M.}
\newblock \bibinfo{title}{Comparing measurements of g< sup>(2)</sup>(0)
  performed with different coincidence detection techniques}.
\newblock \emph{\bibinfo{journal}{JOSA B}} \textbf{\bibinfo{volume}{24}},
  \bibinfo{pages}{2972--2978} (\bibinfo{year}{2007}).

\bibitem{divochiy2008superconducting}
\bibinfo{author}{Divochiy, A.} \emph{et~al.}
\newblock \bibinfo{title}{Superconducting nanowire photon-number-resolving
  detector at telecommunication wavelengths}.
\newblock \emph{\bibinfo{journal}{Nat. Phot.}} \textbf{\bibinfo{volume}{2}},
  \bibinfo{pages}{302--306} (\bibinfo{year}{2008}).

\bibitem{hadfield2009single}
\bibinfo{author}{Hadfield, R.~H.}
\newblock \bibinfo{title}{Single-photon detectors for optical quantum
  information applications}.
\newblock \emph{\bibinfo{journal}{Nat. Phot.}} \textbf{\bibinfo{volume}{3}},
  \bibinfo{pages}{696--705} (\bibinfo{year}{2009}).

\bibitem{migdall2002}
\bibinfo{author}{Migdall, A.~L.}, \bibinfo{author}{Branning, D.} \&
  \bibinfo{author}{Castelletto, S.}
\newblock \bibinfo{title}{Tailoring single-photon and multiphoton probabilities
  of a single-photon on-demand source}.
\newblock \emph{\bibinfo{journal}{Phys. Rev. A}} \textbf{\bibinfo{volume}{66}},
  \bibinfo{pages}{053805} (\bibinfo{year}{2002}).

\bibitem{ma2011multiplex}
\bibinfo{author}{Ma, X.-S.}, \bibinfo{author}{Zotter, S.},
  \bibinfo{author}{Kofler, J.}, \bibinfo{author}{Jennewein, T.} \&
  \bibinfo{author}{Zeilinger, A.}
\newblock \bibinfo{title}{Experimental generation of single photons via active
  multiplexing}.
\newblock \emph{\bibinfo{journal}{Phys. Rev. A}} \textbf{\bibinfo{volume}{83}},
  \bibinfo{pages}{043814} (\bibinfo{year}{2011}).

\bibitem{dolgaleva2011AlGaAs}
\bibinfo{author}{Dolgaleva, K.}, \bibinfo{author}{Ng, W.~C.},
  \bibinfo{author}{Qian, L.} \& \bibinfo{author}{Aitchison, J.~S.}
\newblock \bibinfo{title}{Compact highly-nonlinear algaas waveguides for
  efficient wavelength conversion}.
\newblock \emph{\bibinfo{journal}{Opt. Express}} \textbf{\bibinfo{volume}{19}},
  \bibinfo{pages}{12440--12455} (\bibinfo{year}{2011}).

\bibitem{levy2009nitride}
\bibinfo{author}{Levy, J.~S.} \emph{et~al.}
\newblock \bibinfo{title}{Cmos-compatible multiple-wavelength oscillator for
  on-chip optical interconnects}.
\newblock \emph{\bibinfo{journal}{Nat. Phot.}} \textbf{\bibinfo{volume}{4}},
  \bibinfo{pages}{37--40} (\bibinfo{year}{2009}).

\bibitem{helt2011quantum}
\bibinfo{author}{Helt, L.~G.} \emph{et~al.}
\newblock \bibinfo{title}{Quantum optics of spontaneous four-wave mixing in a
  silicon nitride microring resonator}.
\newblock In \emph{\bibinfo{booktitle}{Quantum Electronics and Laser Science
  Conference}} (\bibinfo{organization}{Optical Society of America},
  \bibinfo{year}{2011}).

\bibitem{collinsCLEOEurope}
\bibinfo{author}{Collins, M.} \emph{et~al.}
\newblock \bibinfo{title}{Spatial multiplexing of monolithic silicon heralded
  single photon sources}.
\newblock In \emph{\bibinfo{booktitle}{CLEO: Europe. Munich, Germany}}, vol.
  \bibinfo{volume}{IB-1.3} (\bibinfo{organization}{Optical Society of America},
  \bibinfo{year}{2013}).

\bibitem{mower2011timeMux}
\bibinfo{author}{Mower, J.} \& \bibinfo{author}{Englund, D.}
\newblock \bibinfo{title}{Efficient generation of single and entangled photons
  on a silicon photonic integrated chip}.
\newblock \emph{\bibinfo{journal}{Phys. Rev. A}} \textbf{\bibinfo{volume}{84}},
  \bibinfo{pages}{052326} (\bibinfo{year}{2011}).

\bibitem{li2008systematic}
\bibinfo{author}{Li, J.}, \bibinfo{author}{White, T.~P.},
  \bibinfo{author}{O{'}Faolain, L.}, \bibinfo{author}{Gomez-Iglesias, A.} \&
  \bibinfo{author}{Krauss, T.~F.}
\newblock \bibinfo{title}{Systematic design of flat band slow light in photonic
  crystal waveguides}.
\newblock \emph{\bibinfo{journal}{Opt. Express}} \textbf{\bibinfo{volume}{16}},
  \bibinfo{pages}{6227--6232} (\bibinfo{year}{2008}).

\bibitem{colman2012shift}
\bibinfo{author}{Colman, P.}, \bibinfo{author}{Combri{\'e}, S.},
  \bibinfo{author}{Lehoucq, G.} \& \bibinfo{author}{{De Rossi}, A.}
\newblock \bibinfo{title}{Control of dispersion in photonic crystal waveguides
  using group symmetry theory}.
\newblock \emph{\bibinfo{journal}{Opt. Express}} \textbf{\bibinfo{volume}{20}},
  \bibinfo{pages}{13108--13114} (\bibinfo{year}{2012}).

\end{thebibliography}

\end{document}